\documentclass[aps,prc,twocolumn,groupedaddress,showpacs]{revtex4}
\usepackage{natbib}
\usepackage{graphicx}
\usepackage{epsfig}
\usepackage{amsmath}
\usepackage{color}
\usepackage{subfigure}

\begin{document}

\title{Soft nuclear equation-of-state from heavy-ion data and implications for compact stars}
\author{Irina Sagert$^1$, Laura Tolos$^2$, Debarati Chatterjee$^3$, J\"urgen Schaffner-Bielich$^3$, and Christian Sturm$^{4,5}$}
\affiliation{
$^1$Department of Physics and Astronomy, Michigan State University, East Lansing, Michigan, 48824, USA\\
$^2$Instituto de Ciencias del Espacio (IEEC/CSIC) Campus Universitat Aut\`onoma de Barcelona, Facultat de Ci\`encies, Torre C5, E-08193 Bellaterra (Barcelona), Spain\\
$^3$Institut f\"ur Theoretische Physik, Ruprecht-Karls-Universit\"at, Philosophenweg 16, D-69120 Heidelberg, Germany\\
$^4$GSI Helmholtzzentrum f\"ur Schwerionenforschung GmbH, Planckstra\ss{}e~1, D-64291 Darmstadt, Germany\\
$^5$Institut f\"ur Kernphysik, J. W. Goethe Universit\"at, Max von Laue-Stra\ss{}e~1, D-60438 Frankfurt am Main, Germany}
\date{\today}
\begin{abstract}
Measurements of kaon production at subthreshold energies in heavy-ion collisions point to a soft nuclear equation-of-state for densities up to 2-3 times nuclear matter saturation density. We apply these results to study the implications on compact star properties, especially in the context of the recent measurement of the two solar mass pulsar PSR J1614-2230. The implications are two-fold: Firstly, the heavy-ion results constrain nuclear matter at densities relevant to light neutron stars. Hence, a radius measurement could provide information about the density dependence of the symmetry energy which is a crucial quantity in nuclear physics. Secondly, the information on the nucleon potential obtained from the analysis of the heavy-ion data can be combined with restrictions from causality on the nuclear equation-of-state. From this we can derive a limit for the highest allowed compact star mass of three solar masses.
\end{abstract}
%%%%%%%%%%%%%%%%%%%%%%%%%%%%%%%%%%%%%%%%%%%%%%%%%%%%%%%%%%%%%%%%%%%%%%
%%%%%%%%%%%%%%%%%%%%%%%%%%%%%%%%%%%%%%%%%%%%%%%%%%%%%%%%%%%%%%%%%%%%%%
\maketitle
\section{Introduction}
%%%%%%%%%%%%%%%%%%%%%%%%%%%%%%%%%%%%%%%%%%%%%%%%%%%%%%%%%%%%%%%%%%%%%%
Recently, the measurement of the Shapiro delay for the millisecond pulsar PSR J1614-2230 provided a new reliable limit for the highest known pulsar mass of $(1.97 \pm 0.04)\:$M$_\odot$ \cite{Demorest10}. Though, also higher neutron star masses up to $(2.74\pm 0.2)\:$M$_\odot$ for PSR J1748-2021 are under discussion \cite{Freire08,Freire08b}, these are not fully obtained from the observation of effects from general relativity and are therefore less reliable. The value for the maximum neutron star mass is of major interest for nuclear physics as it is tightly connected to the properties of dense nuclear matter, especially the stiffness of the nuclear matter equation-of-state (EoS). 
%%%%%%%%%%%%%%%%%%%%%%%%%%%%%%%%%%%%%%%%%%%%%%%%%%%%%%%%%%%%%%%%%%%%%%
\newline
The exploration of nuclear matter at high density and temperature is also a topic of large experimental effort. High energy nucleus-nucleus collisions at collider facilities like RHIC at Brookhaven and the LHC at CERN explore the nature of strongly interacting matter at large temperatures, while the FAIR facility at GSI, Darmstadt, and the NICA facility at Dubna will focus on matter at high baryon densities. 
%%%%%%%%%%%%%%%%%%%%%%%%%%%%%%%%%%%%%%%%%%%%%%%%%%%%%%%%%%%%%%%%%%%%%%
\newline
%%%%%%%%%%%%%%%%%%%%%%%%%%%%%%%%%%%%%%%%%%%%%%%%%%%%%%%%%%%%%%%%%%%%%%
%%%%%%%%%%%%%%%%%%%%%%%%%%%%%%%%%%%%%%%%%%%%%%%%%%%%%%%%%%%%%%%%%%%%%%
Today, firm information exists about the properties of nuclear matter around its saturation density $n_0$. The compression modulus $K_0 = 9 \frac{dp}{dn} \left|_{n_0} \right.$ describes the stiffness of isospin symmetric nuclear matter at $n_0 \sim 0.17\:$fm$^{-3}$. It has been determined to $(235 \pm 14)\:$MeV from the study of giant monopole resonance oscillations of heavy nuclei \cite{Youngblood99}. The second important quantity is the symmetry energy $S(n)$. Ground state masses of nuclei, the neutron skin thickness of heavy nuclei, and measurements of giant dipole resonances point to $S(n_0) = S_0 \sim (28-32)\:$MeV \cite{Li08,Tsang09}. Isospin diffusion data from intermediate energy heavy-ion collisions provides a constraint on the density dependence of the symmetry energy around $n \sim n_0$ \cite{Li06}. The properties of nuclear matter beyond $n_0$ can be extracted from collective flow of nucleons in non-central nucleus-nucleus collisions \cite{Danielewicz02}. However, by selecting non-central collisions, the probed density enhancement is less pronounced. Furthermore, the flow analysis with respect to the compression modulus of nuclear matter is not yet conclusive \cite{Danielewicz02} and the results obtained by different transport models do not agree \cite{Andronic05}.
\newline
\newline
The focus of this work lies on the impact of a soft nuclear matter EoS, as obtained from the KaoS experiment at GSI, Darmstadt, for baryon densities up to $(2-3)\:n_0$ on the properties of compact stars.  We make use of results on K$^+$ meson production in nuclear collisions at subthreshold energies, wherein the probability to produce K$^+$ is highly dependent on the reached particle density in the collision zone, most pronounced at the lowest bombarding energies. The level of compression is in turn controlled by the stiffness of nuclear matter through the nucleon potential U$_N$. The more attractive U$_N$ is, the higher is the produced  K$^+$  meson abundance, which therefore can serve as a probe for the stiffness of nuclear matter \cite{AichKo} . 
\newline
This paper is organized as follows. First, we review the results on K$^+$ meson production from the KaoS experiment and the need of a soft EoS to explain the observed data. In the next section, we study the relevance of the KaoS results for low mass neutron stars and their connection to the nuclear symmetry energy. This question arises inevitably since light neutron stars have low interior densities which can be in the same range as the ones probed by KaoS. As will be discussed in the following, the knowledge of the isospin symmetric nuclear EoS from the KaoS experiment could enable a direct study of the symmetry energy from light neutron stars. The last part of the paper is focused on the determination of the upper limit on neutron star maximum masses via the \textit{highest possible mass}. This can be deduced applying the stiffest causal EoS starting from a fiducial density which is set by the reliability of the nuclear EoS for densities below the critical value. As the highest possible neutron star mass is anti-proportional to the critical density, a large value of the latter provides stronger restrictions. Previous publications discuss values for the highest possible mass assuming critical densities beyond $n_0$. However, as we will argue, our study is the first one to use a nuclear EoS which is experimentally obtained at supra-saturation densities and thereby allows the choice of a fiducial density larger than the saturation density. We conclude the paper with a discussion of the obtained limits for the highest possible neutron star mass with respect to neutron star observations, such as the PSR J1748-2021 pulsar. 
%%%%%%%%%%%%%%%%%%%%%%%%%%%%%%%%%%%%%%%%%%%%%%%%%%%%%%%%%%%%%%%%%%%%%%
%%%%%%%%%%%%%%%%%%%%%%%%%%%%%%%%%%%%%%%%%%%%%%%%%%%%%%%%%%%%%%%%%%%%%%
\section{A soft nuclear equation-of-state from subthreshold K$^+$ meson production in A$+$A collisions}

The measurements on K$^+$ meson production in nuclear collisions at subthreshold energies were performed with the Kaon Spectrometer (KaoS) at GSI. The beam energy dependence of the K$^+$ multiplicity ratio from Au+Au and C+C collisions $(M/A)_{\mathrm {Au+Au}}/(M/A)_{ \mathrm {C+C}}$ at subthreshold energies of 0.8 to $1.5\:$GeV per nucleon has been introduced as a sensitive and robust probe for the stiffness of nuclear matter \cite{Sturm01}.
\newline
%%%%%%%%%%%%%%%%%%%%%%%%%%%%%%%%%%%%%%%%%%%%%%%%%%%%%%%%%%%%%%%%%%%%%%
%%%%%%%%%%%%%%%%%%%%%%%%%%%%%%%%%%%%%%%%%%%%%%%%%%%%%%%%%%%%%%%%%%%%%%
To describe the experimental results, IQMD and RQMD (Isospin and Relativistic Quantum Molecular Dynamics) transport model calculations were performed \cite{Fuchs01,Hartnack06}, applying a Skyrme type $U_N$ with two- and three-body forces. Two parameter sets were chosen so as to reproduce two nucleon potentials with different levels of repulsion at supra-saturation densities. The transport calculations consistently demonstrate that the beam energy dependence of the kaon multiplicity ratio is described best by the nucleon potential with small repulsion. When applied to infinite isospin symmetric nuclear matter the latter corresponds to a soft nuclear equation of state characterized by a stiffness parameter $K=9 \frac{dp}{dn} \left|_{n_0} \right. = 200\:$MeV \cite{Fuchs01}. While the KaoS measurements seem to suggest even lower values of $K \lesssim 200\:$MeV \cite{Fuchs01}, it should also be noted that the isospin asymmetry of the colliding Au+Au system could result in an increase of $K$ by $\sim 10$\% \cite{Piekarewicz02}. 
\newline
%%%%%%%%%%%%%%%%%%%%%%%%%%%%%%%%%%%%%%%%%%%%%%%%%%%%%%%%%%%%%%%%%%%%%%
%%%%%%%%%%%%%%%%%%%%%%%%%%%%%%%%%%%%%%%%%%%%%%%%%%%%%%%%%%%%%%%%%%%%%%
Despite being defined in the same way, the stiffness parameter $K$ and the compression modulus $K_0$ must be clearly distinguished from each other. Since the K$^+$ mesons in the Au+Au system are found to be produced predominantly at supra-saturation densities of $n \sim (2-3)\:n_0$ \cite{Fuchs01,Fuchs06,Hartnack06,Hartnack11}, the KaoS measurements probe nuclear matter in this density range \cite{Hartnack11,Feng}. As a consequence, the stiffness parameter $K\lesssim 200\:$MeV characterizes the nuclear EoS for $n \sim (2-3)\: n_0$ while the compression modulus $K_0 = (235 \pm 14)\:$MeV is determined by the properties of nuclear matter at $n_0$.
%%%%%%%%%%%%%%%%%%%%%%%%%%%%%%%%%%%%%%%%%%%%%%%%%%%%%%%%%%%%%%%%%%%%%%
\newline
The measured attractive nucleon potential and the corresponding soft nuclear EoS \cite{Sturm01,Fuchs01,Hartnack06} have to be tested on their compatibility with neutron star observations. Establishing a link between nuclear properties obtained from experiments and neutron star observations is a non-trivial task \cite{Li97, Klaehn06, Klaehn11}. Nevertheless, in this work we want to investigate the implications of the results from the KaoS collaboration on compact stars. To avoid any assumptions about the unknown high density nuclear matter, we focus on neutron star properties which can be directly linked to the soft EoS at $n \sim (2-3)n_0$. The first chosen observables are the radii and moments of inertia of light neutron stars, which will be discussed in the following. In the last section of the paper, we focus on the limit for the highest allowed compact star mass \cite{Rhoades74}. 
%%%%%%%%%%%%%%%%%%%%%%%%%%%%%%%%%%%%%%%%%%%%%%%%%%%%%%%%%%%%%%%%%%%%%%
\section{Implications of the nuclear symmetry energy on radii and moments of inertia of low mass neutron stars }
%%%%%%%%%%%%%%%%%%%%%%%%%%%%%%%%%%%%%%%%%%%%%%%%%%%%%%%%%%%%%%%%%%%%%%
With their immense interest in nuclear physics \cite{Ferini}, neutron star radii are a focus of major attention \cite{Lattimer07,Ozel10,Steiner10,Suleimanov10,Horowitz03, Bauswein:2011tp,Bauswein:2012ya,Steiner:2012xt}. Limits on radii are obtained from X-ray bursters and low-mass X-ray binaries (see \cite{Steiner10,Steiner:2012xt, Ozel10, Suleimanov11, Zamfir12, Ozel12, Guillot11, Sala12, Hambaryan11} and references therein) whereas such studies could be substantially refined in the future with the Large Observatory For X-ray Timing (LOFT) \cite{Mignani12}. Recently, Bauswein and Janka \cite{Bauswein:2011tp} have studied the tight correlation between the frequency peak of the postmerger gravitational-wave emission with radius measurements, while Guillemot et al. \cite{Guillemot12} have presented a new technique to deduce limits on the moment of inertia of gamma-ray pulsars. 
%%%%%%%%%%%%%%%%%%%%%%%%%%%%%%%%%%%%%%%%%%%%%%%%%%%%%%%%%%%%%%%%%%%%%%
%%%%%%%%%%%%%%%%%%%%%%%%%%%%%%%%%%%%%%%%%%%%%%%%%%%%%%%%%%%%%%%%%%%%%%
\newline
Neutron star radii and moments of inertia are tightly connected to the symmetry energy \cite{Li08b}, very similar to the nuclear symmetry energy dependence of the neutron skin thickness of heavy nuclei \cite{horoprl}. For our study light neutron stars are of special interest due to their low central densities. If the latter are in the same range as the one probed by the KaoS experiment, information on the stiffness of nuclear matter in the neutron star interior can be taken directly from the KaoS data. Being the remaining uncertainty of the nuclear EoS, the symmetry energy can then be extracted by radius and moment of inertia measurements of light neutron stars.
%%%%%%%%%%%%%%%%%%%%%%%%%%%%%%%%%%%%%%%%%%%%%%%%%%%%%%%%%%%%%%%%%%%%%%
%%%%%%%%%%%%%%%%%%%%%%%%%%%%%%%%%%%%%%%%%%%%%%%%%%%%%%%%%%%%%%%%%%%%%%
\newline
For a consistent analysis, neutron star properties and the KaoS data should both be described by the same underlying EoS. We choose a Skyrme type EoS with a nucleon potential similar to the one which was used in the analysis of the KaoS data. The energy per baryon is written as \cite{Prakash88}:
\begin{eqnarray}
&&\frac{E}{A}= m_n\left(1-Y_p\right)+m_p Y_p+E_0 u^{\frac{2}{3}}+B \frac{u}{2}+D\frac{u^{\sigma}}{(\sigma+1)}
\nonumber\\
&+&\left(1-2Y_p\right)^2 \left[ \left(2^{\frac{2}{3}}-1\right)E_0\left(u^{\frac{2}{3}}-F(u) \right)+S_0 u^\gamma\right],
\label{pheneos}
\end{eqnarray}
whereas $u=n / n_0$ is the baryon number density and $E_0$ is the average kinetic energy of nuclear matter at $n_0$ and a proton fraction of $Y_p=0.5$. Two- and three-body forces, described by the terms $B$ and $D$, together with $\sigma$, are fitted to reproduce the binding energy per baryon $E(n_0,Y_p=0.5)=-16\:$MeV, the stiffness parameter $K$, and the saturation density $n_0$. While $S_0$ is varied between $28\:$MeV and $32\:$MeV, its density dependence is chosen as a power law with $u^{\gamma}$ where $\gamma=0.5-1.1$. The values for $S_0$ and $\gamma$ are motivated by heavy-ion experiments and recent analysis of neutron stars in X-ray binaries \cite{Li06,Tsang09,Steiner10,Lattimer12}. Eq.~(\ref{pheneos}) is used to describe matter in the neutron star core, while an inner and outer crust are incorporated following \cite{Negele73,Ruester06}. 
%%%%%%%%%%%%%%%%%%%%%%%%%%%%%%%%%%%%%%%%%%%%%%%%%%%%%%%%%%%%%%%%%%%%%%
%%%%%%%%%%%%%%%%%%%%%%%%%%%%%%%%%%%%%%%%%%%%%%%%%%%%%%%%%%%%%%%%%%%%%%
\newline
We first study neutron stars with $M=1.25\:$M$_\odot$, in accord with the lightest pulsar masses deduced so far from observations \cite{LatPrakBook}.
\begin{figure}
\subfigure{
    \includegraphics[width=7.5cm]{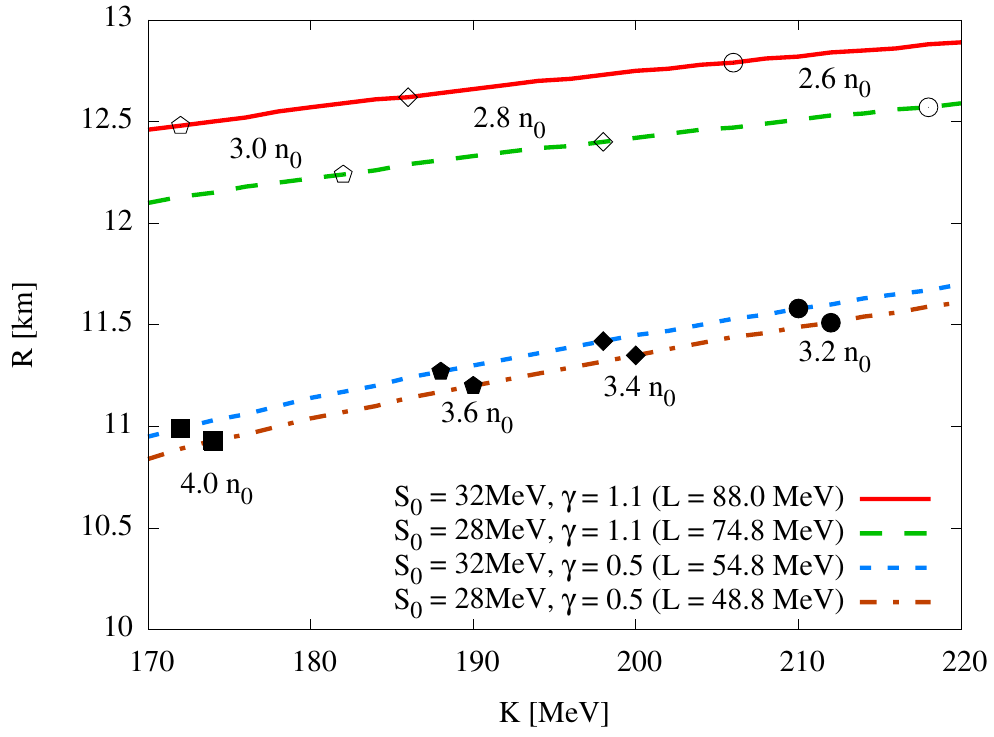}}\hfill
\subfigure{
    \includegraphics[width=7.5cm]{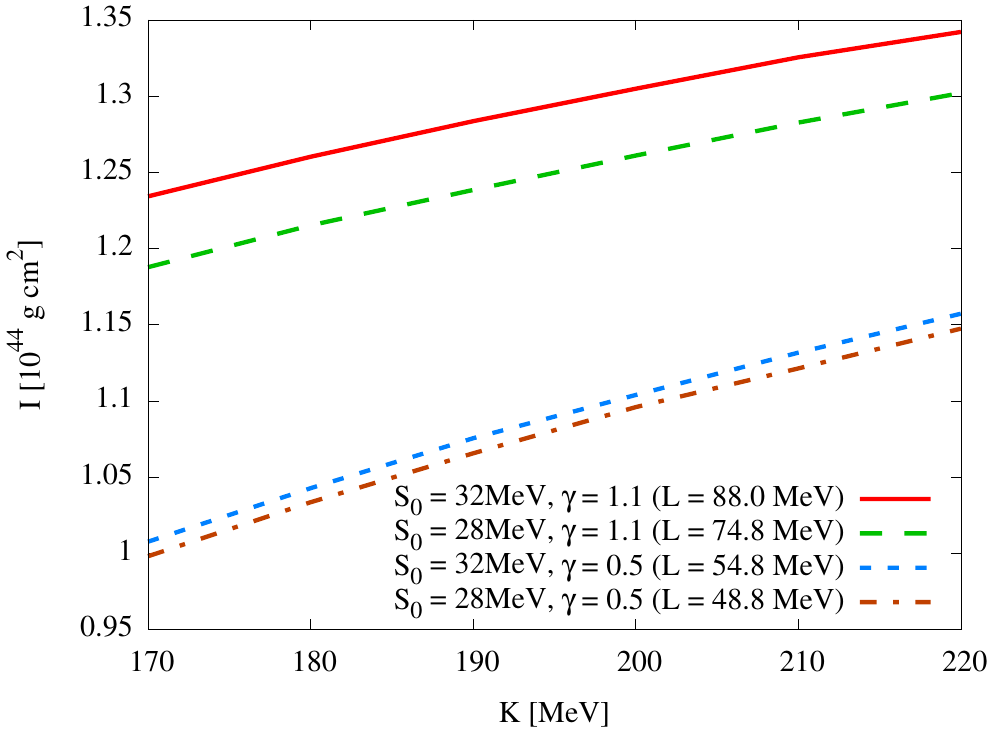}}    
\caption{Radii $R$ and moments of inertia $I$ of non-rotating neutron star configurations with $1.25\:$M$_\odot$ in dependence of $K$ for different setups of $S_0$ and $\gamma$ from Eq.~(1), i.e. different slopes of the symmetry energy $L$. Central densities are given in units of the nuclear matter saturation density $n_0$.}
\label{cenden12}
\end{figure}
The radii $R$ and moments of inertia $I$ are calculated for non-rotating neutron stars, whereas their dependence on $K$ is shown in Fig.~\ref{cenden12} for different symmetry energy configurations. It can be seen that both largely depend on the density dependence of the symmetry energy in form of $\gamma$, i.e. $L=3 n_0 (d S(n_b) / d n_b)| n_0$. At $K \sim 200\:$MeV, stiff and soft symmetry energy configurations lead to a difference in the neutron star radius and moment of inertia of up to $\Delta R \sim 1.5\:$km and $\Delta I \sim  2.5 \cdot 10^{43}  \:$g cm$^2$, respectively. The central densities of the corresponding stars are in the range of $\lesssim 3.4\:n_0$, similar to the density region explored by the KaoS collaboration. At these densities, hyperons, kaons, and quarks, if they appear at all, should not play a dominant role \cite{GM1}. 
\begin{figure}
\subfigure{
    \includegraphics[width=7.5cm]{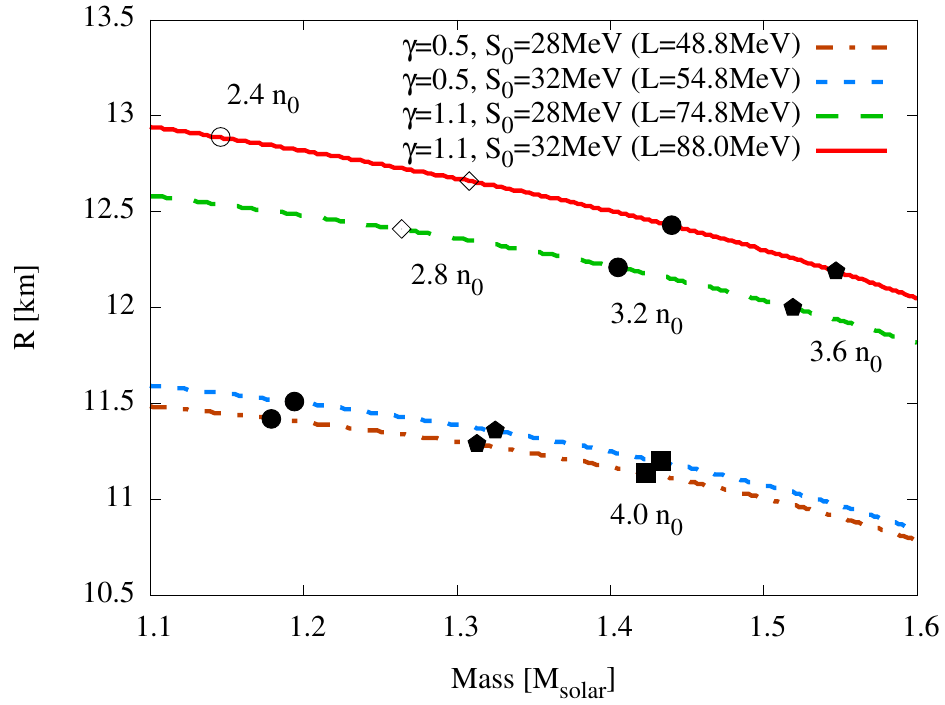}}\hfill
\subfigure{
    \includegraphics[width=7.5cm]{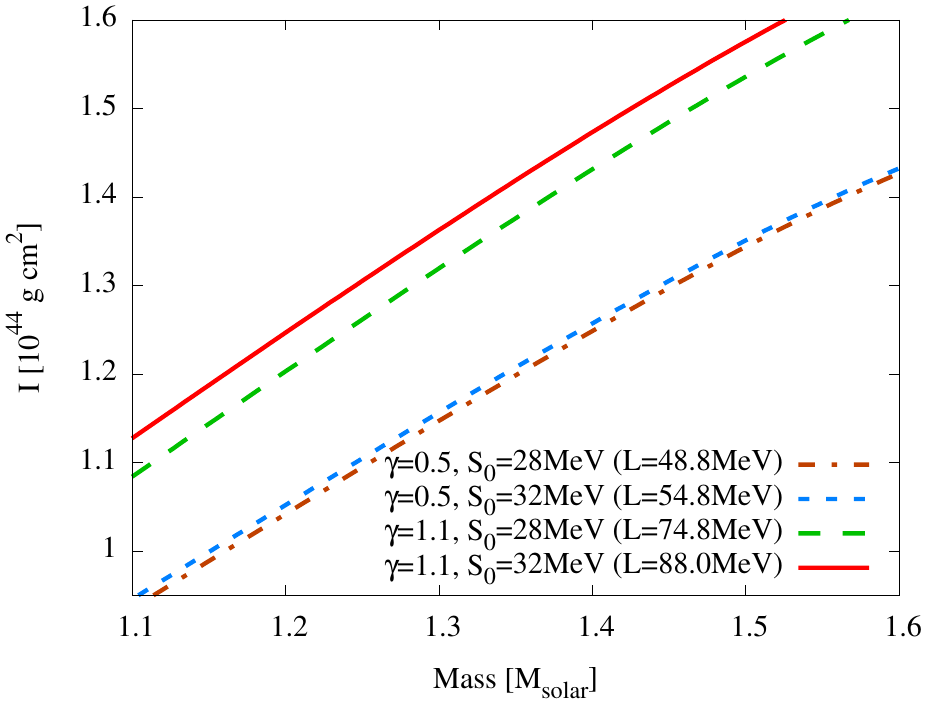}}    
\caption{Radii and moments of inertia of light neutron stars for $K_0 \sim 200\:$MeV in dependence of their mass for different symmetry energy setups as in Fig.~\ref{cenden12}. Central densities are given in units of the nuclear matter saturation density $n_0$.}
\label{moi}
\end{figure}
%%%%%%%%%%%%%%%%%%%%%%%%%%%%%%%%%%%%%%%%%%%%%%%%%%%%%%%%%%%%%%%%%%%%%%
%%%%%%%%%%%%%%%%%%%%%%%%%%%%%%%%%%%%%%%%%%%%%%%%%%%%%%%%%%%%%%%%%%%%%%
However, their possible appearance in a first order phase transition could be probed in the neutrino signal of a galactic supernova explosion \cite{SagertFischer,Dasgupta10}. 
\newline
In Fig.~\ref{moi}, we study the radii and moments of inertia of light neutron stars with masses in the range of $(1.1 - 1.6)\:$M$_\odot$ for $K_0 \sim 200\:$MeV and with equal configuration for the symmetry energy as in Fig.~\ref{cenden12}. It can be seen that neutron stars with masses up to $M \lesssim  1.3\:$M$_\odot$ have central densities which are low enough for the entire neutron star interior to be described by an EoS which is probed by the KaoS experiment. With the isospin symmetric part of the nuclear EoS determined by the latter, radius and moment of inertia measurements from such stars could thereby distinguish between a soft and a stiff behavior of the symmetry energy. 
%%%%%%%%%%%%%%%%%%%%%%%%%%%%%%%%%%%%%%%%%%%%%%%%%%%%%%%%%%%%%%%%%%%%%%
%%%%%%%%%%%%%%%%%%%%%%%%%%%%%%%%%%%%%%%%%%%%%%%%%%%%%%%%%%%%%%%%%%%%%%
\newline
A high precision in neutron star radius measurements could be achieved for compact stars in accreting binaries by future observatories such as the proposed LOFT \cite{Mignani12}. The latter has the promising potential to resolve neutron star radii with a 5\% accuracy and would thereby be able to distinguish between a stiff and soft density dependence of the symmetry energy. For neutron star moments of inertia, only few suggestions exist to deduce informations from stars in specific systems \cite{Lattimer05}. Guillemot et al. \cite{Guillemot12} suggest a new method to obtain a lower limit on $I$ of gamma-ray pulsars using the spin-down of the star coupled to the gamma-ray efficiency. Though this approach requires a very good knowledge of the star's distance, it seems to be especially interesting for low mass neutron stars. Due to the specific shape of the exclusion curve the latter appears to provide larger restrictions to stars with a low moment of inertia.
%%%%%%%%%%%%%%%%%%%%%%%%%%%%%%%%%%%%%%%%%%%%%%%%%%%%%%%%%%%%%%%%%%%%%%
%%%%%%%%%%%%%%%%%%%%%%%%%%%%%%%%%%%%%%%%%%%%%%%%%%%%%%%%%%%%%%%%%%%%%%
\newline
\newline
It is noteworthy that for this study we use a simple ansatz for the nuclear symmetry energy which is usually taken to study neutron stars and the density dependence of the symmetry energy in heavy ion collisions \cite{Li08}. To make precise predictions of light neutron star radii and moments of inertia it is of course necessary to know the exact form of the nuclear symmetry energy, e.g. also a possible non-monotonic behavior with density \cite{Wiringa88}. In addition, large rotational frequencies can increase the neutron star radius and moment of inertia, whereas the latter can also be affected by superfluidity in the neutron star interior. 
\newline
Nevertheless, our study demonstrates that a direct cross-check between results from heavy ion experiments, such as KaoS, and astrophysical observations, in form of radii and/or moments of inertia of light neutron stars, is possible, since both probe nuclear matter in the same density region. Moreover, with the isospin symmetric part of the nuclear EoS at supra-saturation determined from the KaoS data, we provide predictions concerning the relation between the compactness of light neutron stars and the density dependence of the nuclear symmetry energy. 
%%%%%%%%%%%%%%%%%%%%%%%%%%%%%%%%%%%%%%%%%%%%%%%%%%%%%%%%%%%%%%%%%%%%%%
%%%%%%%%%%%%%%%%%%%%%%%%%%%%%%%%%%%%%%%%%%%%%%%%%%%%%%%%%%%%%%%%%%%%%%
\section{Maximum neutron star masses}
We turn now from low mass stars to the maximally allowed gravitational mass of a compact star. The most massive stable neutron star configuration depends strongly on the stiffness of nuclear matter. The softer the nuclear EoS, the lower is the maximum mass which can be reached before the star becomes unstable against collapse to a black hole. 
\newline
Ref.~\cite{Rhoades74} introduced the idea to use known properties of hadronic matter up to a fiducial density $n_{crit}$. At higher densities, the low density EoS is smoothly connected to the stiffest possible EoS allowed by causality. It is given by 
\begin{eqnarray}
p(\epsilon) = p_{crit} +(\epsilon - \epsilon_{crit}), 
\end{eqnarray}
where $p$ is the pressure and $\epsilon$ the energy density \cite{Zeld61}. Their values $\epsilon_{crit}$ and $p_{crit}$ are taken at the transition density $n_{crit}$ from the known low density EoS to the stiffest one. The causal EoS exerts the highest pres$  $sure to balance the gravity of the neutron star mass and therefore leads to the highest possible maximum mass. The existence of such a stiff EoS beyond $(2-3)\: n_0$ can be associated with the emergence of new physics, such as the appearance of quark matter \cite{AlfordNature}, and could be probed by the CBM experiment at FAIR in the near future \cite{CBM}. Previous studies showed that the highest possible mass can be expressed as \cite{Kalogera96,Wiringa88,Hartle78,LatPrakBook}:
\begin{eqnarray}
M_{high} = 4.1 \: \mathrm{M}_\odot \left(\epsilon_{crit} / \epsilon_0  \right)^{-1/2}.
\label{highmass}
\end{eqnarray}
Hereby, the applied EoSs for hadronic matter were obtained from measurements of nuclei at $n \sim n_0$ and then extrapolated to higher densities \cite{Kalogera96,APR,Hartle78}. We want to point out that for such an approach the reliable fiducial density can only be taken as 
\begin{eqnarray}
\epsilon_{crit} = \epsilon_0 = m_n \cdot n_0 \sim 1.4 \cdot 10^{14} \mathrm{g/cm}^3.
\end{eqnarray}
As can be seen from Eq.~(\ref{highmass}), this results in $M_{high} \sim 4.1\:$M$_\odot$. Since for higher $\epsilon_{crit}$ the nuclear EoS is determined via extrapolation from $\epsilon_0$, the corresponding lower values of $M_{high}$, e.g. $M_{high} \sim 2.9\:$M$_\odot$ for $\epsilon_{crit} = 2\: \epsilon_0$ \cite{Kalogera96,APR,Hartle78}, represent only possible predictions for the highest possible mass, and depend on the validity of the extrapolation.
\begin{figure}
\centering
\includegraphics[width=7.5cm]{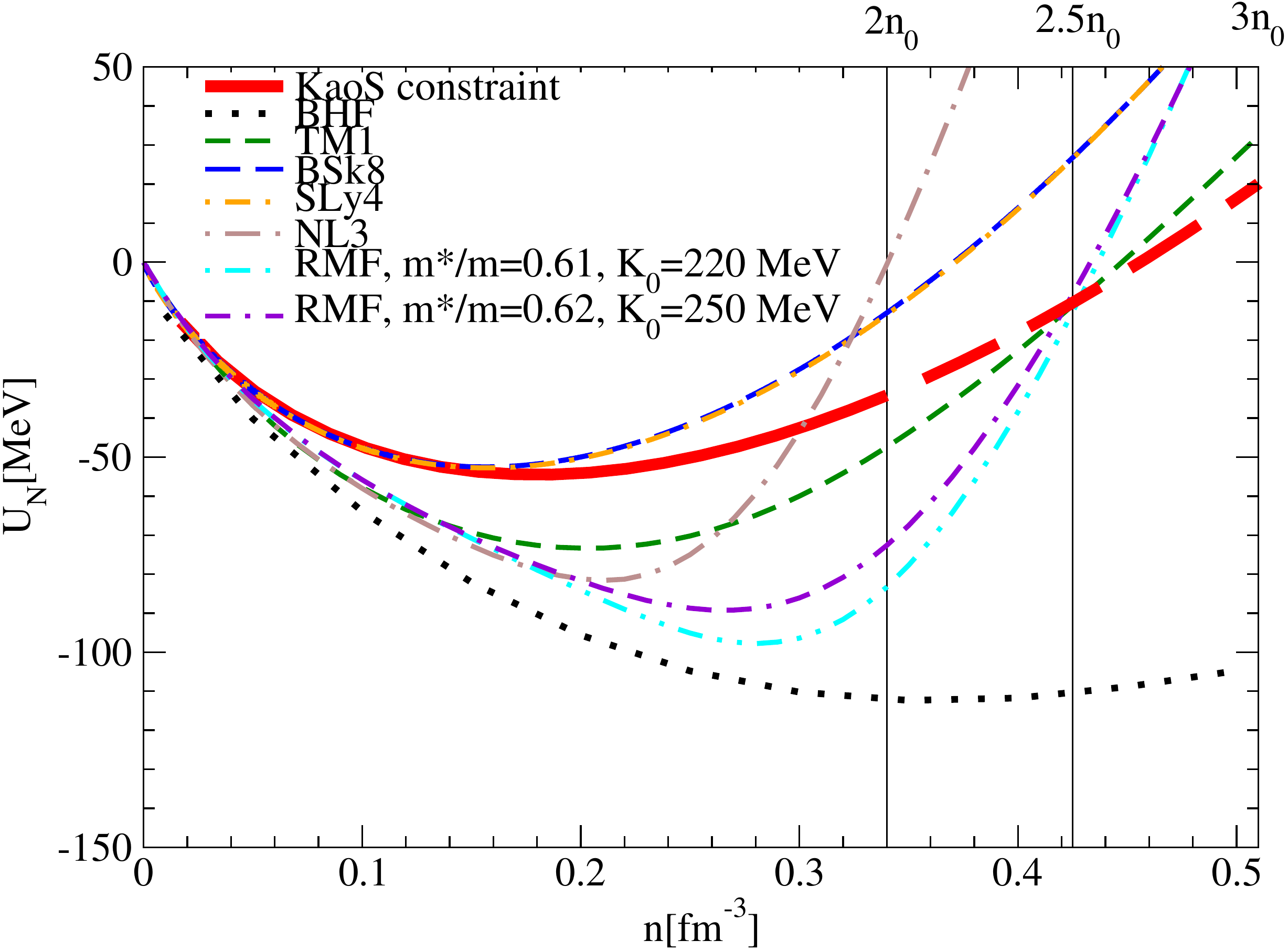}
\caption{The nucleon potential as a function of the baryon number density for different models. The vertical thin lines indicate different densities for the onset of the stiffest EoS.}
\label{fig:un}
\end{figure}
Results from the KaoS collaboration, on the other hand, directly test the EoS for densities $n \sim (2 - 3)\: n_0$ and therefore justify the choice of $n_{crit}$ in the same range. 
\newline
Consequently, in this section we will calculate the highest possible masses with the restriction that for $n_{crit}$ up to $(2-3) \: n_0$ the nucleon potential should fulfill the results from the analysis of the KaoS data. For Skyrme type models this restriction corresponds to $K \leq 200\:$MeV. Today, there exists a large variety of Skyrme parameter sets \cite{Dutra12}. For our work we chose two representative approaches for Skyrme models, BSk8 and SLy4 \cite{Goriely05,Chabanat98}, which are applied in astrophysical and nuclear physics studies. Both models are fitted to reproduce properties of neutron rich nuclei and nuclear matter at saturation density. They have similar values of $K_0 \sim 230\:$MeV while the symmetry energies have a different density dependence. For Relativistic Mean Field (RMF) EoSs, the stiffness of high density nuclear matter is determined by the nucleon effective mass $m^*$ at $n_0$ and not by $K_0$ \cite{BogutaStoecker}. Consequently, the corresponding $U_N$ is chosen by varying $m^*$ for a given $K_0$, so as to obtain a nucleon potential which is similar to or even more attractive than the Skyrme parametrization within the density limits. A more attractive nucleon potential at supra-saturation densities allows a higher compression of matter for the same bombarding energy, which enhances multiple-step processes in subthreshold kaon production and is in line with KaoS results.
\begin{figure}
  \centering
    \includegraphics[width=7.5cm]{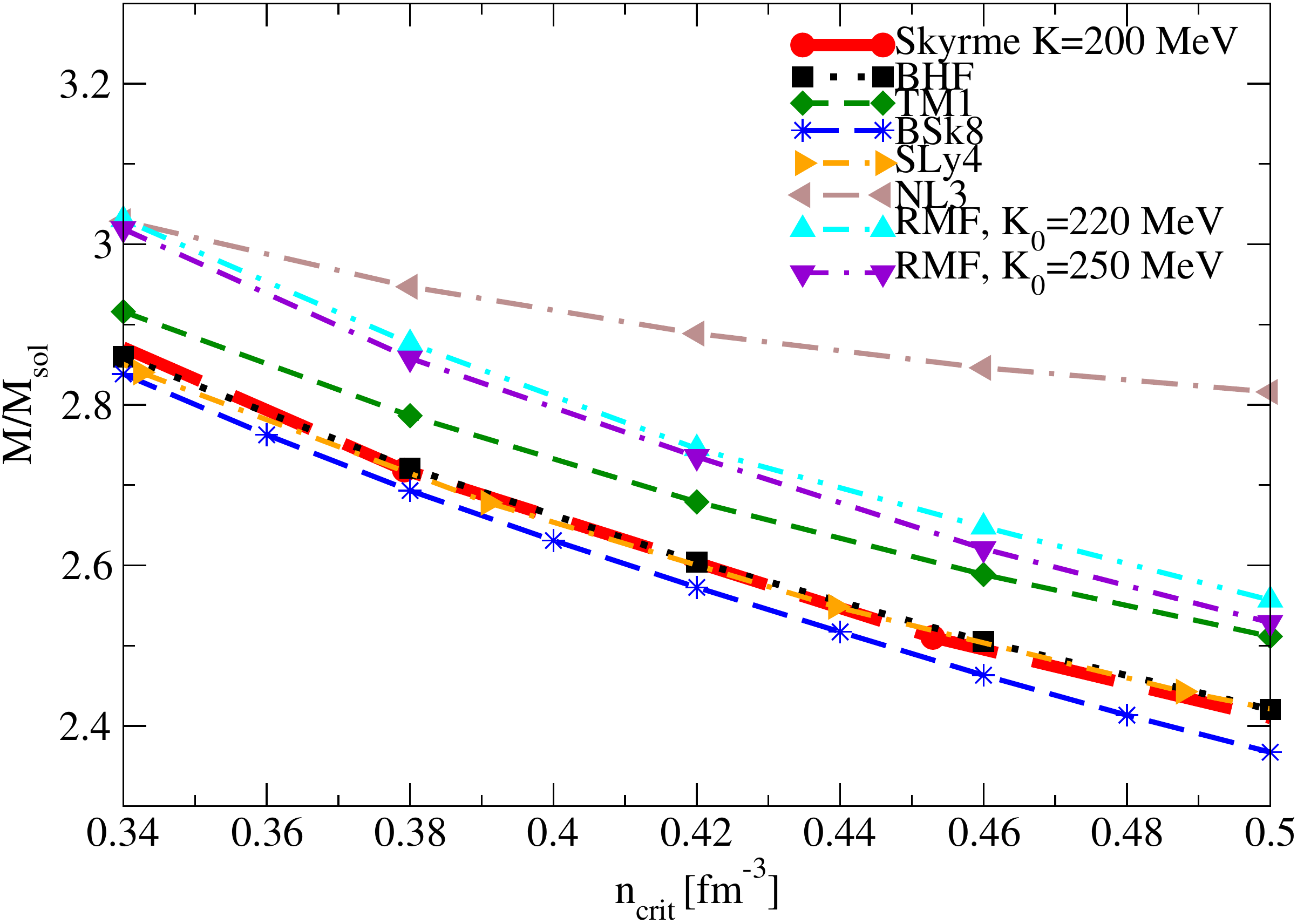}
\caption{Highest possible masses for compact stars according to the approach of \cite{Rhoades74} for different EoS as a function of the critical
density $n_{crit}$ for the onset of the stiffest possible EoS.}
\label{highestm}
\end{figure}
\newline
Fig.~\ref{fig:un} shows the nucleon potentials of the chosen models together with the restriction from the KaoS experiment. Realistic parameter sets should have a $U_N$ below the KaoS constraint up to densities of $n \sim (2-3)\:n_0$. The TM1 \cite{Sugahara94} parametrization as well as the Brueckner-Hartree-Fock approximation (BHF) \cite{isaac,schulze,zuo,bombaci} fulfill this requirement. For a RMF model with $n_0=0.17\:$fm$^{-3}$ and $K_0=220\:$MeV, we arrive at $m^*/m= (0.53-0.65)$ for $n\sim (2-3)\: n_0$ while for $K_0=250\:$MeV we obtain $m^*/m=(0.54-0.67)$ for the same density range. Other schemes, such as BSk8 \cite{Goriely05}, SLy4 \cite{Chabanat98} or NL3 \cite{lalazissis}, produce nucleon potentials which are more repulsive than the Skyrme benchmark in the density region of interest. Therefore, we find that while the BSk8, SLy4, and NL3 parametrizations are fitted to nuclear matter at saturation density they are not applicable for higher values of $n \sim (2-3) n_0$. However, for comparison, we keep these parameter sets in the following calculation.
\newline
Fig.~\ref{highestm} shows the highest possible neutron star masses for $n_{crit} \sim (2-3) \: n_0$ and the discussed EoSs. For $n_{crit} \sim 2 \: n_0$, the star is dominated by the stiffest EoS and therefore reaches masses of up to $3\:$M$_\odot$. Smaller maximum masses  of $M_{high} \sim 2.4\:$M$_\odot$ are obtained for the upper limit of $n_{crit} \simeq 3 \: n_0$. The higher $n_{crit}$ is, the later is the onset of the stiffest EoS in the star's interior. Consequently, less mass is supported and the value for the highest mass decreases. 
%%%%%%%%%%%%%%%%%%%%%%%%%%%%%%%%%%%%%%%%%%%%%%%%%%%%%%%%%%%%%%%%%%%%%%
%%%%%%%%%%%%%%%%%%%%%%%%%%%%%%%%%%%%%%%%%%%%%%%%%%%%%%%%%%%%%%%%%%%%%%
\newline
It can be seen from Fig.~\ref{highestm} that for the Skyrme type EoSs (Skyrme $K=200\:$MeV, BSk8 or SLy4) as well as the BHF calculations the maximum masses are smaller than for the TM1 or NL3 models. While having a nucleon potential well below the KaoS limit, RMF models with varying $m^*$ give a maximum mass which is above the ones for the Skyrme and BHF EoSs. The smaller $m^*$ is, the stiffer is the EoS and, therefore, the larger is the maximum mass. We checked the impact of symmetry energy on our results, varying $S_0$ and $\gamma$ for Eq.(\ref{pheneos}) and the $\rho$ meson coupling constant in the RMF models. Since the maximum mass configuration is dominated by the causal high density EoS and not by asymmetry at low densities, the symmetry energy has very little influence on $M_{high}$, leading to a difference in mass of the order $\Delta M \lesssim 0.02\:$M$_\odot$. As can be seen from Fig.~\ref{highestm}, we can conclude that a pulsar mass of $2.7\:$M$_\odot$ as found by \cite{Freire08} is marginally compatible with a soft EoS and requires a prompt transition from a soft EoS to the stiffest possible at a density around $(2.2 - 2.5)\:n_0$. 
%%%%%%%%%%%%%%%%%%%%%%%%%%%%%%%%%%%%%%%%%%%%%%%%%%%%%%%%%%%%%%%%%%%%%%
%%%%%%%%%%%%%%%%%%%%%%%%%%%%%%%%%%%%%%%%%%%%%%%%%%%%%%%%%%%%%%%%%%%%%%
\section{Summary and Conclusions}
K$^+$ multiplicities from heavy-ion collisions at GSI indicate that the nuclear EoS is soft for densities of $(2 - 3) \: n_0$. To test the consequences on neutron star properties we study low mass neutron stars whose radii and moments of inertia are sensitive to the density dependence of the symmetry energy. We find that light neutron stars with $M \lesssim 1.25\:$M$_\odot$ are strong candidates to have central densities in the range of $\sim 3 \: n_0$ which is within the density region tested by the KaoS collaboration. Such low mass neutron stars are therefore well suited for future radius measurements. 
\newline
Moreover, to test whether a soft nuclear EoS up to $\sim 3 \: n_0$ is compatible with tentatively massive neutron stars such as PSR J1748-2021 \cite{Freire08}, we apply the KaoS results up to densities of $2\: n_0 \leq n_{crit} \leq 3 \: n_0$ \cite{Rhoades74} and introduce the stiffest possible EoS for $n > n_{crit}$ to calculate the highest allowed maximum mass $M_{high}$. Within this approach, the KaoS results confirm the previous theoretical estimation for the highest possible neutron star mass of $\sim 3\:$M$_{\odot}$. A pulsar mass of $2.7\:$M$_{\odot}$ is not excluded by the KaoS data, but requires the onset of the stiffest possible EoS already at $n_{crit} \sim 2.2-2.5 \: n_0$. In the future, the CBM experiment at FAIR will probe densities beyond $3\:n_0$ using rare probes as e.g. D-meson \cite{CBM} and thereby provide further constraints on $M_{high}$.
%%%%%%%%%%%%%%%%%%%%%%%%%%%%%%%%%%%%%%%%%%%%%%%%%%%%%%%%%%%%%%%%%%%%%%
%%%%%%%%%%%%%%%%%%%%%%%%%%%%%%%%%%%%%%%%%%%%%%%%%%%%%%%%%%%%%%%%%%%%%%
\acknowledgments{
We thank Isaac Vida\~na for providing us with the  BHF EoS and Alessandro Brillante for helpful contributions on neutron star properties. Furthermore, we would like to thank Bengt Friman for his comments and suggestions to improve the manuscript. D.C. and I.S. are thankful to the Alexander von Humboldt foundation. I.S. acknowledges the support of the High Performance Computer Center and the Institute for Cyber-Enabled Research at Michigan State University. L.T. is supported from MICINN under FPA2010-16963 and RyC2009, and from  FP7 under PCIG09-GA-2011-291679. J.S.-B. is supported by the DFG through the Heidelberg Graduate School for Fundamental Physics and by the BMBF grant 06HD9127. }

\end{document}